\documentclass[sigconf]{acmart}

\AtBeginDocument{%
  }

\setcopyright{acmlicensed}
\copyrightyear{2023}
\acmYear{2023}
\acmDOI{XXXXXXX.XXXXXXX}

\acmConference[AsiaCCS'24]{ACM ASIA Conference on Computer and Communications Security}{July 2024}{Singapore}
\acmISBN{978-1-4503-XXXX-X/18/06}

\newcommand{\phuzz}{\textsc{Phuzz}}

\usepackage{url}

\usepackage{hyperref}
\usepackage{listings,xcolor}

\definecolor{dkgreen}{rgb}{0,.6,0}
\definecolor{dkblue}{rgb}{0,0,.6}
\definecolor{dkyellow}{cmyk}{0,0,.8,.3}

\usepackage{float}

\newfloat{lstfloat}{htbp}{lop}
\floatname{lstfloat}{Listing}
\lstdefinestyle{php}{
	language        = php,
	basicstyle      = \footnotesize\ttfamily,
	keywordstyle    = \color{dkblue},
	stringstyle     = \color{red},
	identifierstyle = \color{dkgreen},
	commentstyle    = \color{gray},
	emph            =[1]{php},
	emphstyle       =[1]\color{black},
	emph            =[2]{if,and,or,else},
	emphstyle       =[2]\color{dkyellow},
	numbers=left,
	firstnumber=1,
	numberfirstline=true
	numberstyle=\tiny,
	numbersep=-5pt,
	tabsize=1,
	extendedchars=true,
	breaklines=true,
	frame=lines,
	showspaces=false,
	showtabs=false,
	belowskip=0pt,
	showstringspaces=false,
	breakatwhitespace=false,
	showlines=true,
	captionpos=b,
}

\lstdefinestyle{nocoloring}{
	keywordstyle=\color{black},
	commentstyle=\color{black},
	stringstyle=\color{black},
	identifierstyle = \color{black},
	basicstyle      = \small\ttfamily,
	numbers=left,
	firstnumber=1,
	numberfirstline=true
	numberstyle=\tiny,
	numbersep=-4pt,
	tabsize=1,
	extendedchars=true,
	breaklines=true,
	frame=lines,
	showspaces=false,
	showtabs=false,
	belowskip=0pt,
	showstringspaces=false,
	breakatwhitespace=false,
	showlines=true,
	captionpos=b,
}

\begin{document}

\title[\phuzz: A Coverage-guided Fuzzer for PHP Web Applications]{What All the \phuzz~Is About: A Coverage-guided Fuzzer for Finding Vulnerabilities in PHP Web Applications}

\author{Sebastian Neef}
\orcid{0000-0003-3055-0823}
\affiliation{%
  \department{Security in Telecommunications}
  \institution{Technische Universität Berlin}
  \city{Berlin}
  \country{Germany}
}
\email{neef@sect.tu-berlin.de}

\author{Lorenz Kleissner}
\orcid{0009-0005-7341-1366}
\affiliation{%
  \institution{Technische Universität Berlin}
  \city{Berlin}
  \country{Germany}
}
\email{lorenz.kleissner@protonmail.com}

\author{Jean-Pierre Seifert}
\orcid{0000-0002-5372-4825}
\affiliation{%
  \department{Security in Telecommunications}
  \institution{Technische Universität Berlin}
  \city{Berlin}
  \country{Germany}
}
\additionalaffiliation{
  \institution{Fraunhofer SIT}
  \city{Darmstadt}
  \country{Germany}
}
\email{jean-pierre.seifert@tu-berlin.de}
\renewcommand{\shortauthors}{Neef, et al.}

\begin{abstract}
Coverage-guided fuzz testing has received significant attention from the research community, with a strong focus on binary applications, greatly disregarding other targets, such as web applications.
The importance of the World Wide Web in everyone's life cannot be overstated, and to this day, many web applications are developed in PHP. 
In this work, we address the challenges of applying coverage-guided fuzzing to PHP web applications and introduce \phuzz, a modular fuzzing framework for PHP web applications.
\phuzz~uses novel approaches to detect more client-side and server-side vulnerability classes than state-of-the-art related work, including SQL injections, remote command injections, insecure deserialization, path traversal, external entity injection, cross-site scripting, and open redirection.
We evaluate \phuzz~on a diverse set of artificial and real-world web applications with known and unknown vulnerabilities, and compare it against a variety of state-of-the-art fuzzers.
In order to show \phuzz' effectiveness, we fuzz over 1,000 API endpoints of the 115 most popular WordPress plugins, resulting in over 20 security issues and 2 new CVE-IDs.   
Finally, we make the framework publicly available to motivate and encourage further research on web application fuzz testing.
\end{abstract}



\newcommand{\ok}{\textbf{\textcolor{blue}{\checkmark}}}
\newcommand{\nok}{\textcolor{red}{$\times$}}
\newcommand{\timeout}{\textcolor{brown}{$\times$}}

\maketitle

\section{Introduction}\label{sec:introduction}
One programming language, which has been specifically optimized for web development since its inception 25 years ago, is PHP \cite{phpmanual}.
To this day, PHP is the server-side language that drives over 75\% of the websites according to \cite{phpusagew3,serverprogw3}.

Software development mistakes can become security issues affecting the application's confidentiality, integrity, and availability.
Thus, it is in the best interest of development and security teams to identify and remediate any programming mistakes and potential vulnerabilities as early as possible.

Automated security testing tools promise to facilitate the discovery of vulnerabilities.
One such approach is fuzz testing, in which a \textit{fuzzer} generates seemingly random inputs fed to the target application to trigger unintended behavior, potentially leading to vulnerabilities \cite{Li2018}.
Coverage-guided fuzz testing has been an active research field, but many publications (i.e., AFL, AFL++, Hongfuzz, FuzzTest) primarily focus on binary application fuzzing and memory-corruption vulnerabilities \cite{afl_2023,fioraldi2020afl++,honggfuzz_2023,fuzztest_2023,Li2018,Zhu2022}.

These fuzzers are inapplicable to modern web applications written in high-level, interpreted languages such as PHP.
Commercial and free web application fuzzers often follow a black box approach, which limits their effectiveness in detecting vulnerabilities due to no or little feedback from the fuzzed application while relying on pre-defined attack vectors \cite{doupe2010johnny,w3af_2023,burp_2023,acunetix_2023,sqlmap_2023,zap_2023}.

Our work applies coverage-guided fuzz testing to PHP web applications.
We believe that our results have the potential to advance this research field by making the following contributions:

First, we present \phuzz, a framework designed for coverage-guided fuzzing of PHP web applications with the following features:
	\begin{itemize}
		\item A novel crawler-free approach to seeding the fuzzer with endpoints, allowing more fine-grained control over the fuzzing scope.
		\item A novel instrumentation approach without modifications to the fuzzed application's source code or related components (i.e., databases), capable of intercepting PHP exceptions or errors, and bypassing authentication and authorization functionality to collect more code coverage.
		\item A novel vulnerability detection approach to discover more server-side and client-side vulnerability classes than state-of-the-art related work.
		\item Supporting parallel fuzzing out-of-the-box, making it suitable for large-scale fuzzing testing, as our evaluation shows.
	\end{itemize}

Furthermore, we evaluate \phuzz' effectiveness on a set of artificial and real-world web applications and uncover over 20 security issues, including two zero-days, in authenticated APIs.

\section{Background} 
Although the reader is assumed to be familiar with general concepts of the World Wide Web, web security and coverage-guided (binary) fuzz testing, a brief overview of the most critical aspects relevant to our work will be provided in this section.

\subsection{Web Vulnerabilities}\label{sec:webvulns}
The CWE database comprehensively lists over 450 software development mistakes that can lead to a security vulnerability \cite{cwesoftware}.
Although the OWASP Top 10 groups CWEs according to their data \cite{owasptop10}, not all weaknesses are suitable for a fuzz testing approach.
Some weaknesses require domain knowledge, context, and interpretation of the application's behavior to distinguish between legitimate functionality and a vulnerability. 
In other cases, the vulnerability is hard to detect automatically, i.e., business logic flaws or security logging or monitoring failures. 
For this reason, this work focuses on user-supplied input. e.g. HTTP headers, cookies, query, and body parameters, to PHP functions that result in a vulnerability. 

Our \phuzz~framework is capable of detecting the following server-side and client-side vulnerabilities in a target application through coverage-guided fuzz testing:

\subsubsection{Server-side vulnerabilities}
We refer to any vulnerabilities that exploit server-side components of web applications as \textit{server-side} vulnerabilities.

\begin{itemize}
    \item \textit{SQL injection (SQLi)}:\\
    Attacker-controlled input becomes part of a SQL-query sent to a database \cite{owasp_sqli}.
    Example on line 5 of Listing \ref{lst:motivationcode}.

    \item \textit{Command injection (RCE)}:\\
    Attacker-controlled input is executed as part of a shell command \cite{owasp_rce}. 
    Example on line 8 of Listing \ref{lst:motivationcode}.

    \item \textit{Insecure deserialization (IDes):}\\
    Deserialization functions are called with attacker-controlled input \cite{owasp_ides}.
    Example on line 11 of Listing ref \ref{lst:motivationcode}.

    \item \textit{Path traversal (PaTr)}:\\
    The attacker controls parts of the file path passed to file-related functions \cite{owasp_pat}.
    Example on line 14 of Listing \ref{lst:motivationcode}.

    \item \textit{External entity injection (XXE)}:\\
    The attacker can pass XML to vulnerable XML parsers \cite{owasp_xxe}.
    Example on line 18 of Listing \ref{lst:motivationcode}.
\end{itemize}

\subsubsection{Client-side vulnerabilities}
In contrast to server-side vulnerabilities, we refer to any vulnerabilities that target the users of a vulnerable web application in the context of their browser as \textit{client-side}. \phuzz~supports the following vulnerability classes:

\begin{itemize}
    \item \textit{Cross-site scripting (XSS)}:
    The attacker can include HTML markup or JavaScript in a vulnerable website due to missing sanitation \cite{owasp_xss}. There are several variants of XSS vulnerabilities (i.e., reflected, stored, DOM-based), but the malicious payload is executed within a user's browser.
    Example on line 21 of Listing \ref{lst:motivationcode}.

    \item \textit{Open redirection (OpRe)}:\\
    The attacker can manipulate the location to which a browser is redirected \cite{owasp_opre}.
    Example on line 24 of Listing \ref{lst:motivationcode}.
\end{itemize}

\subsection{Fuzzing}
Fuzz testing, also known as fuzzing, is an established and well-researched technique for finding bugs in software \cite{Manes2021}. 
In essence, a fuzzer repeatedly generates test cases, provides them to the application being tested, and monitors its behavior to detect vulnerabilities. 
Fuzzers can be classified into three categories based on the analysis capabilities and access to source code according to \cite{Godefroid2007,Li2018,Zhu2022}:

Whitebox fuzzers have full access to the source code, so they can monitor the internal behavior and states of the fuzzed application and use that information to improve the generated test cases.
Symbolic execution \cite{godefroid2008automated} and tainting \cite{Ganesh2009} are two such approaches.

In contrast, blackbox fuzzers cannot collect information about the target application's internal behavior or state. 
These fuzzers can only monitor the application from the outside and thus have limited ability to optimize test cases. 
In the context of web applications, \cite{doupe2010johnny} discovered that blackbox fuzzers often miss vulnerabilities. 

A strategy in between blackbox and whitebox is called greybox fuzzing.
These fuzzers either work with or without source code while still gathering execution information about the fuzzed application. 
The collected application's execution coverage for a test case becomes the feedback to the fuzzer.
Test cases are prioritized and selected based on the feedback to maximize the observed coverage.
AFL++ is one of many popular state-of-the-art binary application fuzzers employing greybox strategies \cite{fioraldi2020afl++}.

\begin{lstfloat}[t]
\begin{lstlisting}[caption={Example vulnerabilities that \phuzz~is able to find due to its coverage-guidance when initialized with both parameters.}, label={lst:motivationcode},style=php]
	$m = $_GET['m']; $d = $_GET['d'];
	if(substr($m, 0, 1) == "m") {
		if(substr($m, 1, 1) == "s") {
			mysqli_query($db, 
					"SELECT * FROM t WHERE id =  $d"); 
		}
		if(substr($m, 1, 1) == "r") {
			system("echo $d"); 
		}
		if(substr($m, 1, 1) == "u") {
			unserialize($d); 
		}
		if(substr($m, 1, 1) == "f") {
			file_get_contents($d); 
		}
		if(substr($m,1,1) == "e") {
			$doc = new DOMDocument();
			$doc->loadXML($d, LIBXML_NOENT);
		}
		if(substr($m, 1, 1) == "x") {
			echo "$d"; 
		}
		if(substr($m,1,1) == "o") {
			header("Location: " . $d);
		}
	}
\end{lstlisting}
\end{lstfloat}

\subsection{Motivation}
Little work exists on the application of greybox coverage-guided fuzzing to web applications (cf. section \ref{sec:related-work}) due to the research communities' strong focus on binary applications. 
In that research field, it has been shown that greybox fuzzing is effective and suitable for binary applications, and it successfully managed to find several thousand vulnerabilities and bugs in real-world software \cite{ossfuzz_2023}. 

Listing \ref{lst:motivationcode} shows a short PHP script that is vulnerable to all the previously described security issues (cf. section \ref{sec:webvulns}) and simulates how real-world applications might have vulnerabilities buried within nested levels of code. 
A blackbox fuzzer's only possibility to discover the vulnerabilities is to brute-force a correct input for both variables \texttt{\$m} and \texttt{\$d}, which may take a long time and many requests depending on the website's complexity.
With a greybox fuzzing strategy, the fuzzer can use the coverage feedback to optimize the test case generation and eventually find all vulnerabilities. 

\subsection{Challenges}\label{sec:challenges}
When applying the greybox fuzzing technique to web applications, so far primarily used against binary applications, several challenges concerning the instrumentation, test case generation, and vulnerability detection begin to surface. 
Although certain limitations apply, \phuzz~manages to overcome these challenges as described in section \ref{sec:phuzz}. 

\subsubsection{Instrumentation}
Instrumentation is a critical aspect of greybox fuzzing to optimize the generated test cases in order to explore the application's functionality.
Exploring the application's states can be achieved by maximizing the observed code coverage, referred to as coverage-guided fuzzing \cite{Li2018}.
AFL and AFL++ achieve this on binary applications by instrumenting basic blocks of code at compile and sharing the observed coverage over a shared memory region \cite{afl_2023,fioraldi2020afl++}.

This method does not directly apply to web applications implemented in PHP, as the high-level interpreted language is not compiled a priori, but interpreted at runtime in the PHP interpreter. 
Instrumenting the PHP interpreter on a binary application level would primarily detect vulnerabilities in the interpreter itself but not web vulnerabilities in the executed PHP code. 
Therefore, \phuzz~uses the PHP interpreter to instrument the high-level PHP code at runtime.

\begin{figure}[t]
	\includegraphics[width=\linewidth]{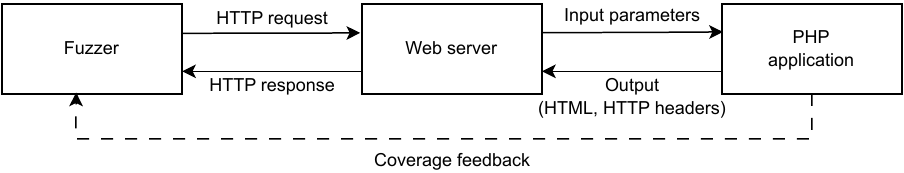}
	\caption{A brief overview of the web application fuzzing setup with the web server between the fuzzer and the PHP application.}
    \Description{A figure depicting how a web server is between the fuzzer (on the left) and the web application (on the right).}
	\label{fig:webfuzz}
\end{figure}

In contrast to binary application fuzzing, where the fuzzer can invoke the fuzzed application directly, there is no direct link between the fuzzer and the fuzzed application due to the web server in the middle, as shown in \autoref{fig:webfuzz}.
The web server receives an HTTP request and passes the relevant input parameters to the PHP application, which returns the information for the HTTP response.
A web server is needed as some web applications rely on its features, i.e., for URL rewriting or access control.

Since HTTP is a network-based protocol and the components might run on separate systems, there are limitations in how coverage information can be shared.
The binary-fuzzing approach of a shared memory region between the fuzzer and the target application is unsuitable, and other solutions must be found to provide the fuzzer with a test case's coverage.
\phuzz~solves this challenge by sharing coverage information using a shared filesystem.

\begin{figure*}[ht!]
	\centering
	\includegraphics[height=0.23\textheight]{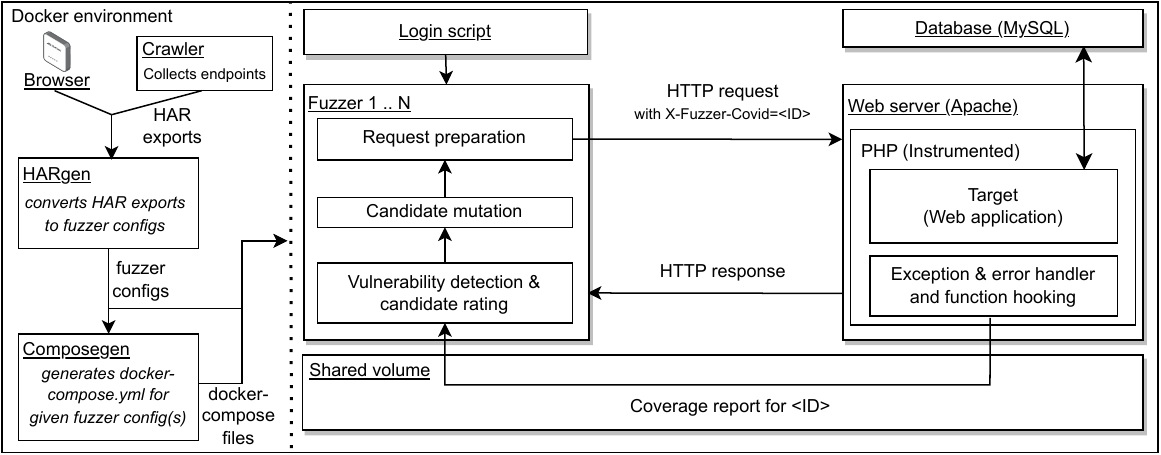}
	\caption{Overview of components involved in the \phuzz~framework.}
    \Description{A figure showing the different components of \phuzz~and their relationship to each other.}
	\label{fig:fuzzcomponents}
\end{figure*}

\subsubsection{Vulnerability detection}
In traditional binary application fuzzing, a crash of the target application is an indicator that the generated test case resulted in a potential vulnerability or unintended behavior.
Again, this is not directly applicable to web application fuzzing since a crash of the PHP interpreter is masked by the web server, which would return a valid HTTP response with one of the 5xx HTTP status codes indicating an error.
Also, errors at the PHP code level will usually not crash the interpreter but result in PHP emitting exceptions or error messages and aborting the execution \cite{php_errors}, again resulting in an error page generated by the web server.

An essential difference between binary and web application fuzzing is the detectable class of vulnerabilities. 
While memory corruption-related vulnerabilities are prevalent in binary application fuzzing, web vulnerabilities are caused by the insecure use of web language-specific (i.e., PHP) functions, thus covering a different set of vulnerability classes  (cf. \cite{owasptop10}). 

Further, PHP offers exception and error-handling mechanisms, allowing applications to handle problems gracefully.
This becomes a challenge for a web fuzzer since vulnerability-related errors (i.e., database errors when SQL injection occurs) can be masked by the application and thus become hard to detect. 
Also, some vulnerability classes (i.e., cross-site scripting) require an analysis of the returned website to be detected, forcing the web fuzzer to wait for the web application to finish and to retrieve the full HTTP response, which can have an impact on the fuzzer's performance.

\phuzz~implements several techniques to detect a multitude of server-side and client-side web vulnerabilities, even in cases where the web application tries to handle errors gracefully.

Further, web applications may use multi-step actions for specific functionality, i.e., guiding a user through multiple forms. 
For this, we differentiate between two scenarios: (1) \emph{independent actions}, i.e., user input on endpoint A, which is then insecurely used at endpoint B, and (2) \emph{dependent actions}, i.e., a unique ID derived form user input on endpoint A, which is then \emph{required} for a consecutive action at endpoint B. 
\phuzz~supports fuzzing of \textit{independent} multi-step actions by configuring a fuzzer instance for every involved endpoint.
Parallel execution of these separate fuzzer instances will discover such multi-step vulnerabilities, as the fuzzer instance for endpoint A provides the inputs, and the fuzzer instances for endpoint B trigger the vulnerability.
Sharing information between fuzzer instances, which is required for scenario (2), is considered future work. 
Nonetheless, support for scenario (1) enables \phuzz~to discover vulnerabilities in multi-step actions.

\subsubsection{Test case generation}
Another set of challenges stems from the test case generation. 
Fuzzers can employ different strategies, including random, mutation-based, or generation-based changes \cite{Li2018,Zhu2022}.
All these methods are suitable for binary application fuzzing, since no inherent structure is required for the inputs to reach the target application.

With web fuzzing, the HTTP requests are required to adhere to the specific structure of the HTTP protocol (cf. appendix Listing \ref{lst:examplehttpreq}), placing certain constraints on the mutation process, i.e., HTTP headers, query, or body parameters, which are limited to specific, allowed characters or encodings \cite{nielsen_hypertext_1999}. 
Also, some web vulnerabilities require the input to have a specific pattern or structure to be triggered and detected, i.e., parameter names or values.

Randomly changing the input on a bit-level will result in HTTP requests that never reach the PHP application due to the web server's rejection for not following the HTTP specification.
While this can also happen to mutation-based strategies, the repeated iteration and selection of best-performing test cases will eventually lead to valid HTTP requests.

Thus, creating suitable test cases must be adapted for web application fuzzers. 
\phuzz~approaches this challenging task by performing byte-level mutations on the input parameters and not the entire HTTP request, thus ensuring that the request is valid and the mutated input reaches the PHP application.

\section{{\phuzz}}\label{sec:phuzz}
\subsection{Components}
\phuzz~ consists of several components enabling a streamlined fuzzing process. \autoref{fig:fuzzcomponents} visualizes the relationship between all components. 
Except for the \emph{Browser}, all components are implemented as Docker containers (\cite{docker}), thus easing future work and reproducibility.

\subsubsection{Browser \& Crawler}
The process begins with obtaining endpoints that are to be fuzzed. We define an \emph{endpoint} as a URL path, and request method with a specific set parameters (HTTP headers, query, body, and cookie parameters).
Similar to related work (i.e., \cite{trickel2022toss,van2021webfuzz}), we also implement a crawling component to discover fuzzable endpoints automatically.
The crawler is based  on a headless chromium browser instrumented by the playwright library \cite{playwright}, enabling it to render and interact with complex web pages, although the current implementation is limited to HTML form submissions.
Using our novel function hooking approach (cf. section \ref{sec:functionhooking}), we can disable anti-crawling protections within the target application. 
However, crawling does not provide much control over the identified endpoints, limiting the application's functionality that will be fuzzed. 
Thus, we implement an additional manual approach: A browser's DevTools are used to capture and save all HTTP requests performed while interacting with the web application. 
Both approaches eventually export their captured endpoints as an HTTP Archive Format (HAR) file, which is passed to the \emph{HARgen}.

\subsubsection{HARgen}
This component generates configuration files for \phuzz~from a given HAR file. 
It applies several filters, i.e., excluding static resources, as finding vulnerabilities these is improbable.
Further, for parameterized endpoints, each parameter can be marked as \emph{fixed} or \emph{fuzz} to allow the fuzzer to keep specific values unchanged (i.e., session identifiers) during the mutation process.
\subsubsection{Composegen}
In order to orchestrate all components and their respective Docker containers to launch \phuzz~with the desired configuration in a reproducible manner, we use the docker-compose extension \cite{docker_compose}. 
\textit{Composegen} generates a \texttt{docker-compose.yml} file from a set of fuzzer configuration files and (pre-)defined configuration options. 

\subsubsection{Login script}
To support fuzzing of endpoints that require authentication or authorization, a login script can be executed during \phuzz'~startup to obtain session cookies, which will be used in all subsequent fuzzing requests to the target.
Another method supported by \phuzz~is hooking into the target's specific authentication and authorization functions to disable them, as explained in section \ref{sec:functionhooking}.

\subsubsection{Fuzzer}
The \emph{Fuzzer} component is the centerpiece of the \phuzz~framework.
It takes the endpoints to be fuzzed from the loaded configuration files and generates test cases, which we call \emph{candidates}.
The mutation-based candidate generation follows a basic energy-based algorithm (\cite{Li2018,Zhu2022}) and is described in more detail in the section \ref{sec:testcasegeneration}.
The new candidate and its parameters are then converted to a valid HTTP request in the \emph{Request preparation} step, which is then sent to the web server with a specific \texttt{ID}.
After receiving an HTTP response, the fuzzer evaluates the associated coverage and debug information.
This information is then processed by multiple vulnerability detectors to identify security issues and rate the candidate.
The candidate with the highest rating is chosen for the next mutation iteration.
Multiple instances of the fuzzer can run in parallel and synchronize generated test cases, similar to popular binary fuzzers like AFL.
All critical algorithms, such as candidate mutation and vulnerability detection, are interchangeable to facilitate future work. 

\subsubsection{Shared volume}
\phuzz~uses an in-memory Docker volume to provide a shared filesystem for information exchange between all containers.
This allows the feedback information collected on the web server to be shared with the fuzzer for evaluation.
\subsubsection{Web server}
As \phuzz~only instruments the PHP interpreter, it is agnostic to the web server, which is Apache by default.
A dedicated web server allows the processing of multiple HTTP requests in parallel and thus increases \phuzz' performance.

\subsubsection{Database}
Web applications may require an SQL database or other external components to work.
While \phuzz' instrumentation does not require changes to the database or other external components, it comes with a MySQL database by default.

\subsection{Instrumentation}
Naturally, our coverage-guided approach requires collecting coverage information for the fuzzed application.
\phuzz' instrumentation has several benefits compared to related work, as its instrumentation does not require any modification of the target's PHP code, web server, database, or other components (cf. \cite{trickel2022toss,van2021webfuzz,Zhao2022}).
In fact, the instrumentation is transparent to the fuzzed application, as it only happens on the PHP interpreter level, thus minimizing the risk of negatively affecting the application's or other components' functionality.

\subsubsection{Coverage collection}
\phuzz~supports two coverage collection extensions for PHP: Xdebug (\cite{xdebug_coverage}) and PCOV (\cite{pcov_coverage}).
While PCOV claims to have significantly better performance than Xdebug, \phuzz~additionally allows to constrain the coverage collection to specific directories, as more tracked files can further affect the overall performance. 

Since the web server can handle multiple requests simultaneously, the fuzzer generates a unique \emph{ID} (\texttt{<timestamp>-<UUIDv4>}) per candidate, which is included in the request as an additional \texttt{X-Fuzzer-Covid} header.
An optional \texttt{X-}header reduces the risk of interfering with the target, as it does not affect the length of the URL, body, or cookie parameters, which otherwise might exceed allowed length limits.
Our PHP-side instrumentation code saves the collected coverage information as a file named \texttt{<ID>.json} on the shared volume.
After receiving the HTTP response from the web server, the fuzzer can evaluate a candidate's coverage and debug information by reading the corresponding file.

In contrast to related work, our approach of separating the HTTP response and the coverage information is sensible for the following reasons: 
It does not change the response's content; thus, there is less risk of interfering with or breaking the target's functionality.
Second, the original response can be passed to client-side vulnerability detection modules, while server-side vulnerability detection modules can analyze the coverage and debug information.

Although the coverage collection is transparent to the target, the instrumentation's start and stop functions need to be executed \emph{before} and \emph{after} the target's PHP code.
The former can be achieved using PHP's \texttt{auto\_prepend\_file} configuration option \cite{php_ini}.
Similarly, \texttt{auto\_append\_file} runs a file after the actual PHP code terminates, except when \texttt{exit()} is called \cite{php_ini}.
To circumvent this limitation, \phuzz' configures a \texttt{register\_shutdown\_function}, which executes upon script termination, including due to \texttt{exit()} or reaching the maximum execution time.
A target cannot interfere since shutdown functions are called in the order they were registered, and the instrumentation is the first code to run \cite{php_register_shutdown_function}.

However, PHP applications can terminate unexpectedly, for example, due to uncaught errors or exceptions.
Nonetheless, these cases can be interesting for finding vulnerabilities.
Thus, \phuzz~registers custom exception and error handlers with PHP's built-ins to save the coverage information when an unhandled event occurs. 
In order to prevent the target application from overriding \phuzz' custom error and exception handlers, we use \emph{function hooking} to ensure \phuzz' handlers are always called. 

\subsubsection{Function hooking}\label{sec:functionhooking}
UOPZ (User Operations for Zend) is a PHP extension that provides functionality to manipulate PHP functions on the interpreter level \cite{php_uopz}.
\phuzz~makes extensive use of the \texttt{uopz\_set\_return} feature to instrument native and non-native PHP functions. 
Listing \ref{lst:uopzhook} shows a simplified example in which the function \texttt{func} is overridden by a function that calls the original \texttt{func}, with all arguments passed through, and returns the original \texttt{func}'s return value.

Hooking into PHP functions for instrumentation has several benefits for \phuzz:
First, it captures function calls and their arguments, allowing \phuzz~to identify function calls where attacker-controlled input is passed to potentially dangerous functions, which helps confirm vulnerability findings generated by \phuzz.
Second, it catches exceptions and errors emitted by the original function, which can be an indicator of malformed input passed to the function and, thus, potential vulnerabilities.
Next, it ensures that \phuzz' error and exception handlers are always set and cannot be changed by the target.
Last, it enables overriding authentication, authorization, access control, bot protection, anti-crawling or other specific functions of a target to fuzz otherwise unreachable code paths.
\phuzz~uses all these features to detect vulnerabilities, as detailed in the following section.

\begin{lstfloat}[t]
\begin{lstlisting}[language=php,caption={Instrumenting native and non-native PHP functions with UOPZ.},label=lst:uopzhook,style=php]
	uopz_set_return('func', function($args) {
		try {
			$ret = func($args);
		} catch($e) { /* Exception handling */ }
		// Log function name, arguments, errors and exceptions
		return $ret;
	}, true);
\end{lstlisting}
\end{lstfloat}

\subsection{Vulnerability Detection}\label{sec:vulndetection}
\phuzz~is capable of detecting a multitude of server-side and client-side vulnerabilities, as listed in section \ref{sec:webvulns}.
\subsubsection{Server-side vulnerabilities}
The function hooking approach enables \phuzz~to detect many server-side vulnerabilities. 
\phuzz~hooks native PHP functions, which are highly likely to be used and eventually be called by the target, regardless of the number of abstraction layers (i.e., web frameworks).
Further, additional debug information (i.e., function name, arguments, error message) is collected when an exception or error is raised due to invalid input. 
The collected exceptions and errors are exposed to the \emph{Fuzzer} component, whose \emph{VulnChecker} can then inspect and decide to raise a vulnerability alert or not.
\phuzz~can effortlessly be extended with additional function hooks to support other vulnerability classes or technologies, i.e., SQLite or MSSQL.
In the appendix, Table \ref{tab:phuzz-uopz-hooks} lists all hooked PHP functions for each of the following vulnerability classes:

\paragraph{SQL injection (SQLi)}
    \phuzz~hooks into four functions used by our targets to query MySQL databases (cf. section \ref{sec:evaluation}).
    Fuzzer-controlled input will eventually result in malformed SQL statements passed to these functions, resulting in a detectable exception or error.

\paragraph{Command injection (RCE)}
    \phuzz~identifies these vulnerabilities by hooking four functions that will pass the input to the system's shell (i.e., \texttt{/bin/sh}) and checking for error messages indicating invalid commands or arguments, i.e., syntax errors, unknown commands, or invalid file paths.

\paragraph{Path traversal (PaTr)}
    \phuzz~hooks a total of 48 functions that take a filename or path as an argument to perform file or directory-related operations to discover path traversal vulnerabilities.

\paragraph{Insecure deserialization (IDes)}
    Invalid input to the \texttt{unserialize()} function emits an error detected by \phuzz~to identify potential PHP object injection vulnerabilities.

\paragraph{External entity injection (XXE)}
    Two functions of PHP's built-in XML parser are hooked to detect the processing of invalid XML entities in a vulnerable configuration (\texttt{LIBXML\_NOENT}).

\subsubsection{Client-side vulnerabilities}

\phuzz~can detect two classes of client-side vulnerabilities.
Detecting these does not require server-side instrumentation but just an unmodified HTTP response from the target to be analyzed by the fuzzer, which we achieve using our instrumentation strategy.

\paragraph{Cross-site scripting (XSS)}
\phuzz~implements the XSS detection algorithm proposed by webFuzz \cite{van2021webfuzz}. 
In essence, a specific marker is inserted during the candidate mutation and later checked for existence in the DOM of the returned page, enabling \phuzz~to detect reflected and persistent XSS vulnerabilities. 

\paragraph{Open redirection (OpRe)}
Similar to XSS, open redirection vulnerabilities are detected by analyzing the HTTP response.
\phuzz~checks the status code for the redirection range (300-399) and the redirection URL provided by the \texttt{Location} header for fuzzer-controlled input.

\subsubsection{Validation method}
\phuzz~will generate many candidates, some of which will eventually contain malformed input to any hooked function, thus causing a failure (i.e., exception, error, \texttt{false} return value) detectable by \phuzz' instrumentation.
However, not all failures are automatically a vulnerability.
For example, a check for the non-existence of a file could be legitimate functionality, but \texttt{file\_exists} returning \texttt{false} would be classified as a potential bug and thus coverage and debug information saved by the PHP-side instrumentation.

Therefore, the fuzzer component's task is to determine whether the saved coverage and debug information is a potential vulnerability.
For each vulnerability class, the \emph{VulnCheck} interface must be implemented with adequate detection rules. 
The \emph{VulnChecker} interface allows grouping of multiple VulnCheck implementations. 
After receiving the HTTP response from the web server, the fuzzer passes the candidate to the configured VulnChecker, which will pass it to all its VulnChecks for analysis.
Each VulnCheck examines the candidate and debug information to trigger a vulnerability alert if the conditions are met.

One VulnChecker that \phuzz~implements is the \emph{ParamBasedVulnChecker}, which aims to minimize the false-positive rate by ensuring that any of the candidate's \textit{fuzz} parameters is part of the arguments passed to a hooked function.
Alternatively, \phuzz~provides a \emph{DefaultVulnChecker} which raises a vulnerability alert when an exception or error in a hooked function was observed, making it prone to false-positives as not all exceptions or errors originate from vulnerable function use.

\subsection{Test Case Generation}\label{sec:testcasegeneration}
Another important part of a fuzzer is its test case generation strategy.
Optimizing mutation algorithms is an extensive research area on its own, and thus, we acknowledge that \phuzz~has only basic strategies, although these are inspired by binary-level fuzzers such as AFL.
We believe that \phuzz' modularity allows future work to improve and implement more advanced mutation strategies.

\subsubsection{Seeds}
Similar to other fuzzers, \phuzz~requires initial seed values and parameter names to generate the initial set of candidates.
The \emph{fuzzer configs} (cf. \autoref{fig:fuzzcomponents} and the example provided in the appendix as Listing \ref{lst:fuzzconfig}) provide all information to form the HTTP request to the target, including the endpoint's URL, HTTP request method, HTTP headers, query parameters, body parameters, and cookie parameters.
A parameter can be defined to be \emph{fixed} (does not change) or \emph{fuzz} (will be mutated), thus making \phuzz~highly flexible.
If multiple parameters or values for a parameter are provided, the fuzzer generates combinations in order to maximize the coverage.

\subsubsection{Scoring \& selection}
After the initial generation, all candidates are evaluated against the target to obtain a base score.
\phuzz~provides a \texttt{ScoringFormula} interface and a default implementation, which rates a candidate based on a simple metric: The number of new paths and lines executed server-side.
Following the energy-based mutation pattern (\cite{Li2018}), \phuzz~always selects the next candidate $c$ with the highest score, calculates energy $E$, and mutates this candidate $E$ times. 
Each new candidate $c'$ is evaluated against the target, and the scoring and selection process repeats.
Should all $E$ mutations $c'$ of a candidate $c$ \emph{not} find new paths, the next lower candidate is chosen as $c$.
This approach favors candidates that find new paths and thus more coverage in the fuzzed application while ensuring that all candidates will eventually be mutated to maximize the chance of finding new vulnerabilities.

\subsubsection{Mutation}
In order to find new coverage and vulnerabilities in the target, a candidate's parameters need to be altered, a process usually called \textit{mutation} \cite{Li2018}.
Although web applications can handle binary input, i.e., used for file-uploads, \phuzz~focuses on byte-level mutation of strings, i.e., query and body parameters, HTTP headers, or cookies, as these cover most of the HTTP protocol and have strict limitations on what characters are allowed according to the HTTP standard.
Since PHP converts numerical strings to integers, \phuzz~byte-level modifications impose no limitations.

\phuzz~provides the interfaces \texttt{Mutator} and \texttt{ParamMutator}, which require the function \texttt{mutate} to be implemented.
The former has a list of ParamMutators and returns a list with new mutations for a provided string.
The actual byte-level changes happen in the ParamMutators, which take a string and return a new, mutated version.
\phuzz~implements an initial set of 10 ParamMutators mutate by adding, removing, replacing, or swapping characters or digits at randomly chosen positions, mimicking the functionality of other (binary) fuzzers.

Additionally, \phuzz~implements three special mutators whose task is to insert markers for specific vulnerability classes. In contrast to fuzzers with static payload lists, \phuzz' markers will be mutated as well, thereby generating more test cases: 
The \emph{ProtocolPrefixMutator} prefixes a string with \texttt{http://}, \texttt{https://}, or \texttt{ftp://} to increase the likelihood of finding open redirections.
The \emph{PathTraversalPayloadParamMutator} adds \texttt{../} or \texttt{/etc/passwd} by chance of 5\% anywhere in the string to increase the chance of finding path traversal vulnerabilities.
The \emph{XSSPayloadParamMutator} adds one of three typical XSS payloads with webFuzz' marker anywhere in the string, every 20 mutations on average \cite{van2021webfuzz}.

\subsubsection{Request preparation}
\phuzz' request handling, and thus parameter encoding, is done by the \texttt{python-requests} package (\cite{python_requests}).
Each candidate has several properties (URL, HTTP method and headers, cookies, query, and body parameters) needed to create an HTTP request.
We assume body parameters to be used only in \texttt{POST}, \texttt{PUT}, and \texttt{DELETE} requests, with the special case of encoding the parameters as a JSON dictionary when the \texttt{Content-Type} header is set to \texttt{application/json}.
For all other request methods, i.e., \texttt{GET}, \texttt{OPTIONS}, and \texttt{TRACE}, \phuzz~supports query parameters, cookies, and headers.
The fuzzer generates a HTTP request and sends it to the target with a configurable timeout.
After receiving the response, its main properties (i.e., status code, response headers, and content) are set on the candidate object for vulnerability analysis and scoring.

\subsubsection{Synchronization}
\phuzz~was designed with parallel fuzzing in mind.
Multiple fuzzer instances can run in parallel, and each instance will perform the aforementioned tasks of generating new candidates, sending them to the target, checking for vulnerabilities, and evaluating their scores.
To prevent identical candidates from being generated by different instances, synchronization between all instances is necessary and achieved using the shared volume.
For each candidate, a hash is generated from several of its properties to ensure that candidates generating the same code coverage will not be evaluated twice.
Therefore, the hashes of the already mutated and evaluated candidates are synchronized and checked before choosing a new candidate.

\section{Evaluation}\label{sec:evaluation}
To answer the following research questions, we evaluate \phuzz~and a set of other state-of-the-art vulnerability scanners against artificial and real web applications with known and unknown vulnerabilities.
\begin{itemize}
	\item[RQ1] Does \phuzz' novel approach at instrumentation and vulnerability detection interfere with or negatively impact the fuzzed target?
	\item[RQ2] How effective is \phuzz~at discovering known and unknown vulnerabilities in PHP web applications compared to other fuzzers?
\end{itemize}

\subsection{Known Vulnerabilities}\label{sec:eval:known}
We consider known vulnerabilities to be publicly available ones for which a technical description and exploit payload exists.

\subsubsection{Vulnerability scanners \& web applications}
Similar to related work \cite{atropos,trickel2022toss}, we first establish a ground truth by comparing \phuzz~with a set of state-of-the-art vulnerability scanners or fuzzers against a set of six different web applications with a diverse set of vulnerabilities.

For a direct comparison, the open-source vulnerability scanners Zed Attack Proxy (ZAP) \cite{zap_2023}, Wapiti \cite{wapiti}, WFuzz \cite{wfuzz} and closed-source BurpSuite Professional \cite{burp_2023} were chosen due to their popularity and presence in related work.  
We did not consider webFuzz \cite{van2021webfuzz}, as \phuzz~mirrors the implementation of its XSS detection algorithm, and no other vulnerability classes are supported.
For CeFuzz \cite{Zhao2022} and Atropos \cite{atropos}, we used the results reported by the authors and related work as the source-code to these two fuzzers were unavailable at the time of writing this paper.
As Atropos' authors promised to open source their prototype, we contacted the authors to inquire about the state of their release. We promptly received an answer that the prototype is not yet publishable as it is still being rewritten to ease future work. 
Similarly, no functional prototype of Witcher \cite{trickel2022toss} was available to us for the evaluation due to the operational complexity and academic nature of the code base \cite{witcher_gh_i1,witcher_gh_i14,atropos}, requiring us to rely on related work for results.
While Witcher’s source code is publicly available, it is no longer under active development by the authors and issues left unanswered \cite{witcher_gh_i14,witcher_gh_i1}. Despite great efforts, we faced severe difficulties and challenges running the prototype due to the limited documentation and, in particular, due to dependency and build issues, whose technical details we document in our repository. The Atropos’ authors acknowledged similar difficulties with Witcher, although they eventually obtained some results with the help of Witcher’s authors \cite{atropos}, which we did not have. 
BackREST \cite{gauthier2021backrest} and RESTler \cite{restler}, both API fuzzers, were not considered in this study as closed-source BackREST only supports NodeJS applications and RESTler requires API specifications to derive inputs, which is orthogonal to the nature of the fuzzed targets and our coverage-guided approach.

As the fuzz targets, we chose 6 different web applications commonly used for evaluations by related work:
\begin{itemize}
    \item The \emph{buggy web app} (bWAPP) \cite{bwapp_old} updated for PHP7 \cite{bwapp} provided 30 vulnerabilities usable for our comparison.
    \item The \emph{Damn Vulnerable Web Application} (DVWA) \cite{dvwa} offered 18 suitable vulnerabilities with varying difficulty levels.
    \item The \emph{Xtreme Vulnerable Web Application} (XVWA) \cite{xvwa} implements 10 vulnerabilities, including insecure deserialization.
    \item \emph{WackoPicko} \cite{doupe2010johnny} had 7 suitable vulnerabilities, after being modified for PHP8 compatibility.
    \item 22 real-world \emph{WordPress plugins} with known vulnerabilities.
\end{itemize}

To obtain the 22 WordPress plugins, we took over 6,500 vulnerability reports from a popular WordPress vulnerability database from 2019-2022 and categorized them into the vulnerability classes based on a report's title.  
For each category, we chose up to 5 plugins with an unauthenticated vulnerability whose source code and exploit payload are publicly available, and the exploitation does not require additional information (e.g., nonces) or sequential requests.

In total, we have 87 vulnerabilities covering a diverse set of classes in different web applications to be discovered by \phuzz~and the other fuzzers.
We categorize the vulnerabilities into client-side and server-side depending on the attack surface and detection method, as stated in sections \ref{sec:webvulns} and \ref{sec:vulndetection}.
Vulnerabilities (e.g., reflected, stored, or DOM XSS) that target the web application's users or browser, e.g., by changing the browser's or website's behavior, which can be detected based on the web server's HTTP response, are considered client-side in this work, even if the server processed the input.
On the contrary, vulnerabilities (e.g., SQLi) that target the web application through vulnerable PHP code and that can be detected by \phuzz' instrumentation, e.g., function hooking or exception and error capturing, are considered server-side.

\subsubsection{Fuzzer setup}
We run all experiments on a Linux system with an AMD Ryzen 7 4800H 8C/16T CPU, 64GB RAM, and NVMe storage unless otherwise specified.
Each fuzzer is seeded with the HTTP request (or equivalent command line arguments) to the vulnerable endpoints of the application, including parameters set to \texttt{fuzz}, to shift the focus on vulnerability detection instead of crawling capabilities.
\phuzz~is evaluated with its \texttt{DefaultVulnChecker} and \texttt{ParamBasedVulnChecker} vulnerability detection methods, but all fuzzers are run with default or the minimum required options, and a timeout of $t=300s$ to avoid hangs or indefinite execution.
Using our function-hooking approach, we disable authentication or authorization checks as needed for the fuzzers to reach the vulnerable endpoint.

\subsubsection{Results}

\begin{table}
\small
\caption{Evaluation of \phuzz~and other vulnerability scanners against web applications with known vulnerabilities.}
\label{tab:ground-truth}
\begin{tabular*}{\linewidth}{@{}lllllll@{}}
\toprule
Application ($N$ vuln.)                 				& \phuzz & BurpSuite & ZAP & Wapiti & WFuzz \\ \midrule
\textbf{bWAPP (30)}                 & 100\%    &     93\%      &  77\%   &       70\%                             &   0\%    \\ 
\multicolumn{1}{l}{3x RCE$^\star$}   	&   2+1$^1$    &    2       & 2    &        0                            &   0    \\
\multicolumn{1}{l}{9x SQLi$^\star$}	  &   9    &    9       &    9  &      7                               &  0     \\
\multicolumn{1}{l}{2x PaTr$^\star$}     &    1+1$^1$   &    2       &  2   &      2                               &   0    \\
\multicolumn{1}{l}{14x XSS$^\dag$}   &   14    &   13        &   8  &      10                               &    0   \\
\multicolumn{1}{l}{2x OpRe$^\dag$}  &    2   &       2    &   2  &      2                               &    0   \\ \midrule
\textbf{DVWA (18)}          		  &   100\%&  83\% & 67\%    &  78\%            &  0\%     \\ 
\multicolumn{1}{l}{3x RCE$^\star$}   	 &   3    &  3 &   1  &      3             &   0    \\
\multicolumn{1}{l}{6x SQLi$^\star$} 	&   6    &  4 &   3  &     2           &    0   \\
\multicolumn{1}{l}{3x PaTr$^\star$} 	 	&1+2$^1$ & 2 & 2 &    3           &   0    \\ 
\multicolumn{1}{l}{6x XSS$^\dag$}  	&   6    &  6 &   6  &      6            &    0   \\ \midrule
\textbf{XVWA (10)}               &  100\%     &   60\%        &  50\%   &   50\%                                  &     0\%  \\
\multicolumn{1}{l}{1x RCE$^\star$}   &    1   &    1       &   1  &    1                                 &     0  \\
\multicolumn{1}{l}{4x SQLi$^\star$} &   4    &   2        &  1   &  0                                   &  0     \\
\multicolumn{1}{l}{1x PaTr$^\star$}  &   1$^1$    &    1       &  1   &  1                                   &  0     \\
\multicolumn{1}{l}{1x IDes$^\star$} &  1     &  0         &  0   &  0                                   &    0   \\
\multicolumn{1}{l}{2x XSS$^\dag$}  &   2    &    1       &   1  &  2                                   &   0    \\
\multicolumn{1}{l}{1x OpRe$^\dag$} &    1   &     1      &   1  &        1                             &  0     \\ \midrule
\textbf{WackoPicko (7)}        &  100\%     &  71\%         & 86\%    &        71\%                             &    0\%   \\ 
\multicolumn{1}{l}{1x RCE$^\star$}   &  1    &   0      &  0   &   0                                  &   0    \\ 
\multicolumn{1}{l}{1x SQLi$^\star$} &  1     &  1         & 1    & 0                                    &   0    \\
\multicolumn{1}{l}{1x PaTr$^\star$}  &   1$^1$    &     0      &  1   &   1                                  &    0   \\
\multicolumn{1}{l}{4x XSS$^\dag$}  &   4    &   4        &   4  &        4                             &   0    \\ \midrule
\textbf{WP Plugins (22)}       &  95\%     &   50\%   &  36\%   & 41\%                                    &   0\%    \\ 
\multicolumn{1}{l}{5x SQLi$^\star$} &   5    &    0       &  0   &  0                                   &   0    \\
\multicolumn{1}{l}{5x PaTr$^\star$}  &   4    &   1        & 1    &    1                                 &    0   \\
\multicolumn{1}{l}{2x IDes$^\star$} &  2     &    0       &  0   &     0                                &    0   \\
\multicolumn{1}{l}{5x XSS$^\dag$}  &    5   &      5     &  3    &  3                                   &  0     \\
\multicolumn{1}{l}{5x OpRe$^\dag$} &   5    &    5       &   4  &   5                                  & 0      \\ \midrule 
\textbf{Total (87)}          &   99\%     &    75\%       &   62\%  &         62\%                            &     0\%    \\
\multicolumn{1}{l}{\textbf{$^\star$: Server-side (48)}} & 98\% & 58\% & 52\% & 44\% & 0\% \\
\multicolumn{1}{l}{\textbf{$^\dag$: Client-side (39)}} & 100\% & 95\% & 74\% & 85\% & 0\% \\

\bottomrule
\end{tabular*}
\\ $^1$~Indicated by the captured internal application exceptions and errors.
\end{table}

\begin{table}
	\centering
     \caption{The number of discovered ($V_d$) and existing vulnerabilities ($V_e$) reported for \phuzz, Atropos, Witcher and CeFuzz.}\label{tab:fuzz-related-work}
	\begin{tabular}{lcccc}
        \toprule
		\shortstack{Application\\~} & \shortstack{\phuzz\\ $V_d$/$V_e$} & \shortstack{Atropos\\ $V_d$/$V_e$} & \shortstack{Witcher\\ $V_d$/$V_e$} & \shortstack{CeFuzz\\ $V_d$/$V_e$} \\ \midrule
		  bWAPP & 30/30 & 27$^{1}$/27$^{1}$ & 0$^{1}$/27$^1$ & 5$^{1}$/5$^1$ (100\%$^3$) \\
        DVWA  & 18/18 & 15$^{1}$/16$^{1}$ & 0$^{1}$/16$^1$ & 3$^{1}$/3$^1$ (100\%$^3$) \\
        XVWA  & 10/10 & 7$^{1}$/9$^{1}$ & 3$^{1}$/9$^1$ & - \\
        WackoPicko & 7/7 & - & 2$^2$/3$^2$ & - \\
        \bottomrule
	\end{tabular}
    \\ $^1$ Reported by Atropos \cite{atropos}, $^2$ Reported by Witcher \cite{trickel2022toss}, \\ $^3$ Reported by CeFuzz \cite{Zhao2022}, \texttt{-} no results available
\end{table}

Table \ref{tab:ground-truth} shows the results of the ground-truth evaluation. 
\phuzz~identifies 99\% of the vulnerabilities and thus outperforms the state-of-the-art black-box vulnerability scanners BurpSuite, ZAP, Wapiti, and WFuzz. 
Tables \ref{tab:fuzz-dvwa} - \ref{tab:fuzz-known-vulns-wp} in the appendix provide more detailed results by vulnerability class for each fuzzer.

Analogous to the reports of \cite{atropos,Zhao2022}, WFuzz does not report any vulnerability, which is inherent to its design of only reporting general information about the received HTTP responses by default.
The remaining fuzzers discover the majority of the client-side vulnerabilities which they can detect by analyzing the HTTP responses, but face limitations in detecting server-side vulnerabilities within our experiments.

\phuzz' novel instrumentation and feedback approach outperforms all other fuzzers in both vulnerability categories combined by a margin of 24\%, as indicated by the total percentages in Table \ref{tab:ground-truth}.
Especially in the server-side vulnerability realm, the instrumentation and feedback collection have an advantage over the traditional black-box fuzzing approach, which allows \phuzz~to detect 40\%-54\% more vulnerabilities of this type.
For client-side vulnerabilities, \phuzz~finds up to 26\% more vulnerabilities than the other fuzzers, although BurpSuite finds almost the same amount (95\%).

\phuzz~misses one path traversal vulnerability in one of the WordPress plugins, which neither of the tools discovered due to the required traversal path.
However, with longer execution times, \phuzz~should eventually generate a suitable mutation to report a finding.

\phuzz~finds some path traversal and code execution vulnerabilities based on captured internal application exceptions or errors, as annotated in Table \ref{tab:ground-truth} with ~$^1$, since these vulnerabilities originated in PHP expressions, such as \texttt{eval} or \texttt{include}. 
Although \phuzz~cannot instrument PHP expressions, invalid input reaching these will raise a PHP exception or error, which are captured by \phuzz'~custom exception and error handlers. The captured information then serves as an indicator for the respective vulnerability. Since \emph{VulnCheck}s do not take this information into account yet and we manually reviewed all fuzzers' output to determine the discovered vulnerabilities, we also included exceptions and errors that \phuzz~displayed.

Due to the unavailability of a functional prototype of Atropos, CeFuzz, and Witcher for our evaluation, we limit the comparison to the results reported by the respective authors and related work, as shown in Table \ref{tab:fuzz-related-work}.
The changes in the number of existing vulnerabilities ($V_e$) are likely due to differences in the experimental setup or the supported vulnerability classes. For example, CeFuzz only supports RCE and reports to have found all such vulnerabilities in bWAPP and DVWA, which Atropos reports as five and three respectively.
Similarly, Atropos uses nine vulnerabilities in XVWA, while we select ten vulnerabilities since \phuzz~supports open redirections.
Due to the limited availability of detailed vulnerability information, we only compare the number of successfully identified vulnerabilities as stated by related work.

In summary, by comparing \phuzz~against the state-of-the-art using a common set of well-known vulnerable applications, we established that \phuzz~is 24\%-38\% more effective at uncovering vulnerabilities and it finds more vulnerabilities in each artificial application than related work.

\subsection{Unknown Vulnerabilities}\label{sec:unknownfuzzing}
The second experiment aims to find previously unknown vulnerabilities, often called zero-days, in real-world source code from popular WordPress plugins.

\subsubsection{Obtaining WordPress plugins}
We consider WordPress plugins as a suitable target since they fulfill the following criteria: 
(1) Publicly available, including their source code and (2) high impact due to their immense active installation count.

Except for some commercial plugins, a plugin's source code is open, allowing us to download and install it in our environment.
A plugin can be developed by a third-party individual or entity with varying degrees of programming knowledge, creating the chance of finding insecure code and, thus, a vulnerability while challenging \phuzz' instrumentation against different programming paradigms.

WordPress.org's plugin store, configured by default in all WordPress installations, lists metadata for over 55,000 plugins.
The active installation count varies greatly, so we define a cut-off at a minimum of 300,000 active installations to find vulnerabilities affecting a large number of websites.
Further, less popular plugins might be outdated or incompatible with recent versions of WordPress.
Thus, we downloaded 183 plugins with a total installation count of 180M.
We found 121 plugins with less than 1M active installations and 62 with up to and over 5M.
Therefore, a critical vulnerability in any of the highly popular plugins could have a devastating effect on up to several million websites (i.e., \cite{wp_plug_vuln1,wp_plug_vuln2}). 
\subsubsection{Identifying fuzzable endpoints}
A plugin can offer any functionality and change WordPress in many ways; for example, Woocommerce turns the CMS into an e-commerce website \cite{woocommerce}.
Further, a plugin's functionality might depend on its configuration or the existence of other plugins, i.e., Woocommerce addons.
Thus, due to the sheer amount of targeted plugins, their complex and divers functionality, and better reproducibility, we chose a programmatic approach to identifying fuzzable endpoints.

WordPress plugins can expose API functionality by registering actions with a specific prefix: \texttt{wp\_ajax\_<apiname>} for authenticated and \texttt{wp\_ajax\_nopriv\_<apiname>} for unauthenticated endpoints (cf. example in Listing \ref{lst:wpapi} in the appendix).
To call an API, a request to \texttt{/wp-admin/admin-ajax.php} with a parameter \texttt{action= <apiname>} needs to be sent.

For each downloaded plugin, we extract all its defined API names and the code of the associated handling functions.
Some plugins do not define API endpoints, or we cannot discover them. 
Thus, we find 1,952 API definitions and 1,715 function handlers in 136 plugins, from which we successfully extract over 95\% of the function handlers for further analysis. 

Next, we analyze an API handler's source code to obtain parameters that are used within the function call that can serve as potential attacker-controlled input, provided through \texttt{\$\_REQUEST}, \texttt{\$\_GET}, \texttt{\$\_POST},  \texttt{\$\_COOKIE}.
Again, we omit a few plugins, which may be using different ways to access parameters, i.e., through helper functions.
In total, we extracted 1,058 parameters and 1,090 associated API endpoints, covering a set of 115 WordPress plugins, whose details we provide in our repository.

As we extract both unauthenticated and authenticated endpoints, we use our function hooking approach to disable several security measures provided by WordPress, including nonce validation, referrer validation, authorization and authentication checks.
Based on the collected information, we generate fuzzer configuration and docker-compose files to run \phuzz~against each endpoint.

\subsubsection{Fuzzer setup}
We automate the process of fuzzing all 1,090 endpoints with a Python script that coordinates the execution of \phuzz~and BurpSuite. 
We choose BurpSuite as the only contender, since it is the best-performing vulnerability scanner available to us, according to Table \ref{tab:ground-truth}.
As each endpoint is fuzzed individually, the web application and database are reinitialized to prevent interference before continuing with the next endpoint.

We limit the maximum fuzzing time for each iteration to $t=180s$. 
Both fuzzers run on a server with 8 cores/16 threads (Intel E5520) and 48 GB of RAM. 
Since BurpSuite has multi-core support, \phuzz~was configured to use $N=10$ parallel fuzzer instances, which leaves sufficient resources for the database and web server.

Again, both tools are initialized with identical HTTP requests.
Fuzzable parameters were seeded with the string \emph{fuzz} and the coverage collection path constraint to the plugin's respective directory.

\subsubsection{Results}

\begin{table}
	\centering
	\caption{Findings generated by fuzzing popular WordPress plugins for unknown (0-day) vulnerabilities }\label{tab:phuzz-unknown-vulns-wp}
	\begin{tabular}{lcc}
        \toprule
		{Vuln. class} & \shortstack{{\phuzz}\\(Plugins/APIs/Valid)} & \shortstack{{BurpSuite Pro}\\(Plugins/APIs/Valid)} \\ \midrule
		XSS		&  14 / 24  /  7 &  9 / 16  / 8 \\ 
		PaTr	&  20 {\footnotesize (47)} / 37 {\footnotesize (110)} / 16 & 1 / 1 / 1 \\ 
		SQLi	&  6   / 9  /  0 & 0 / 0 / 0 \\ 
		OpRe   & 1  / 1  /  1 &  1 / 1 / 1\\ 
		\bottomrule
	\end{tabular}
\end{table}

Both \phuzz~and BurpSuite took over 50 hours each to fuzz all 1,090 API endpoints.
BurpSuite generated reports for 98\% of the fuzzed endpoints, resulting in over 1,843 reported issues.
The high number of reported issues stems from BurpSuite's greater variety of tested security issues, including several misconfiguration issues that were raised in every report due to our environment, i.e., the disclosure of private IP addresses.
Disregarding these unrelated issues, BurpSuite has only raised a total of 18 reports matching vulnerability classes mentioned in section \ref{sec:vulndetection}, of which 8 were false positives.

In contrast, \phuzz~generated  691 reports for over 12.5\% of the fuzzed endpoints belonging to 47\% of the fuzzed plugins.
A minor implementation bug in the path traversal detection method has caused a twofold increase in reported API endpoints from 37 to 110 and reported plugins from 20 to 47, which became obvious and effortlessly correctable when reviewing the findings.

An extensive manual review of the reports from both fuzzers was conducted to verify their validity.
\phuzz~discovered 24 valid security issues, more than twice as many valid issues as found by BurpSuite.
In fact, \phuzz~found all vulnerabilities that BurpSuite discovered, except for one XSS.
From Table \ref{tab:phuzz-unknown-vulns-wp} it becomes evident that BurpSuite's strength is detecting client-side vulnerabilities as it lacks feedback from server-side errors and exceptions.
Thus, \phuzz' instrumentation and feedback collection excel at detecting server-side issues, thus letting it identify more server-side issues than BurpSuite.
While both fuzzers discovered cross-site scripting, open redirection, and path traversal issues, \phuzz~was the only one to identify SQL errors.

Albeit these SQL errors were not caused by SQL injections, it showcases how \phuzz~can assist in finding regular programming mistakes, i.e., missing values in \texttt{ORDER BY} clauses, non-existing database tables, or deadlocks.
Both fuzzers fail to take a \texttt{Content-Type: application/json} response header into account when reporting XSS issues, leading to false-positives since no HTML will be rendered by a browser.
This is a known limitation of webFuzz' (\cite{van2021webfuzz}) XSS detection algorithm, which \phuzz~implements.
In the 21 false-positive path traversal cases, \phuzz~detected fuzzer-controlled input to file-related functions, where the input was adequately sanitized or tested for dangerous characters by the application.
While \phuzz~currently cannot infer what transformations were applied to the input before reaching the hooked function, detecting these potential input sinks is beneficial to identify vulnerable data-flows and is thus considered a feature rather than a limitation.

\section{Discussion}

\subsection{Zero-Day Vulnerabilities}\label{sec:zero-day-vulns}
The authors share strong beliefs in responsible disclosure and thus contacted all plugin developers affected by valid findings unearthing previously unknown security issues.

Even though all issues discovered by both fuzzers were exclusively in authenticated APIs, which require either administrative privileges or the knowledge of secrets (i.e., nonces) to be exploitable, almost 60\% of the contacted developers were quick to respond and consider implementing additional in-depth measures.
Thus, multiple issues had been proactively remediated at the time of writing, even though the risk of in-the-wild exploitation is considered low.  

Furthermore, two findings qualified for a CVE-ID under the rules of the Certified Numbering Authority (CNA) WPScan \cite{wpscan_cna}, as these vulnerabilities can be exploited in a multisite WordPress installation, where two classes of administrators exist: super-administrator and subsite-administrator.
In contrast to singlesite WordPress instances or super-administrators, subsite-administrators do not have permission to install plugins or otherwise modify the WordPress instance, i.e., by editing files.
We reviewed our discovered security issues for such boundary violations and submitted them to the WPScan CNA, which acknowledged that two findings were valid and yet unknown vulnerabilities, thus assigning the following CVE-IDs:

\subsubsection{CVE-2023-6294: popup-builder SSRF \& Arbitrary File Read}
The plugin's API endpoint \texttt{sgpb\_import\_subscribers} eventually passes \texttt{importListURL}'s value to \texttt{file()} and \texttt{wp\_remote\_get()}.
The former allows reading any file's content and the latter issuing HTTP requests originating from the vulnerable WordPress instance (Server-Side Request Forgery [SSRF]).
In both cases, the attacker sees the first line of the content, i.e., \texttt{root:x:0:0:root: /root:/bin/bash} when reading \texttt{/etc/passwd}.
The arbitrary file read vulnerability was discovered by \phuzz~and during the review the SSRF was spotted.
The SSRF could have been found by \phuzz, if \texttt{wp\_remote\_get} was hooked and included in the vulnerability detection.

\subsubsection{CVE-2023-6295: so-widgets-bundle Local File Inclusion}
The API endpoint \texttt{so\_widgets\_bundle\_manage} uses the value of the parameter \texttt{widget} to define a widget's path using string concatenation, which is first passed to \texttt{file\_exists()} and then to \texttt{include\_once}.
An attacker can use this to include local files or to potentially gain remote code execution when including a suitable PHP file.

\subsection{RQ1: Impact on Fuzzed Targets}
In contrast to related work, \phuzz' instrumentation does not require changes to the target's or PHP interpreter's source code or any external components.
One instrumentation approach used by related work (e.g., \cite{van2021webfuzz,Zhao2022}) actively changes the fuzzed target's PHP code, which has the potential risk of breaking the application.
To eliminate this risk, \phuzz~does not change the PHP code but hooks into a selected set of PHP functions to run additional code.
Thus, \phuzz'~instrumentation, including the error and exception detection, coverage collection, and the workaround to run the instrumentation before the called PHP script, is almost transparent to the fuzzed target.
In fact, \phuzz~only requires placing the target's source code into the web server's \texttt{DocumentRoot} to make it a fuzz target.  
This reduces friction to set up \phuzz, increasing the likelihood of being used by developers, especially since the tooling around \phuzz~(i.e., HARgen and Composegen) further facilitates configuring and running \phuzz~against a web application.

For all our fuzzed applications, we observed no negative impact or interference of \phuzz' instrumentation on the fuzzed target.
However, should incompatibilities with other fuzz targets occur, each function hook can be individually rewritten or disabled.

\subsection{RQ2: Effectiveness and Limitations of \phuzz}
We have evaluated \phuzz~against state-of-the-art vulnerability scanners (BurpSuite Pro, ZAP, Wapiti, WFuzz) and compared it against related work (Atropos, Witcher, CeFuzz) using a large variety of artificial and real-world vulnerabilities.
Tables \ref{tab:ground-truth}, \ref{tab:fuzz-related-work} and \ref{tab:phuzz-unknown-vulns-wp} show that \phuzz~outperforms its contenders in the amount of successfully discovered vulnerabilities.
Further, we demonstrate \phuzz' applicability and effectiveness for large-scale vulnerability research by fuzzing over 100 of the most popular WordPress plugins in an automated fashion.
We identified over 20 new security issues, including two new zero-days, in authenticated APIs, which require administrator privileges or knowledge of secret tokens to reach the vulnerable code path.
However, the two zero-days are fully exploitable in a multisite WordPress installation.

To the best of our knowledge, \phuzz~is the first greybox fuzzer for PHP to support a diverse set of client-side \emph{and} server-side vulnerability classes, including the baseline from related work of SQL injections, cross-site scriptings, and remote command executions, and raising the bar with support for open redirection and external entity injection vulnerabilities.

While the evaluated blackbox fuzzers find a competitive amount of client-side vulnerabilities as our approach, \phuzz~surpasses them in detecting server-side vulnerabilities due to its instrumentation and coverage-guided approach.
Blackbox fuzzers often use a predefined set of payloads, while greybox fuzzers' coverage-guidance allows them to explore an application and generate mutation-based inputs that trigger a vulnerability.
The effectiveness of greybox fuzzing over blackbox fuzzing has also been observed by related work, e.g., Atropos \cite{atropos} or Witcher \cite{trickel2022toss}.

Naturally, \phuzz~has its own set of limitations, including a few shortcomings regarding vulnerability detection, some of which also apply to related work.
For one, its support for complex vulnerabilities that require multiple actions is limited.
While stateful fuzzing is a hard problem acknowledged by related work, \phuzz~has basic support for multi-step fuzzing of independent actions. 
For example, \phuzz~discovers the second-order SQL injection in DVWA as shown in Table \ref{tab:fuzz-dvwa} in the appendix.
Atropos uses snapshots to discover this stateful vulnerability; an interesting approach that could be adopted into \phuzz~by future work.  

Furthermore, due to limitations of UOPZ and webFuzz, some instances of DOM-based XSS and PHP expression-based vulnerabilities, e.g., path traversal through \texttt{require} or code execution through \texttt{eval}, can only be indirectly detected by the captured errors and exceptions when invalid input reaches these functions.

\subsection{Related Work}\label{sec:related-work}
To the best of our knowledge, we find the following related work focusing on coverage-guided fuzzing of PHP web applications:

\subsubsection{Atropos}
The most recent and concurrent work in this domain is \cite{atropos}, which was only available as a prepublication for USENIX 2024 at the time of writing.
Atropos is a promising coverage-guided fuzzer for PHP web applications with several differences and some similarities to \phuzz.
One major difference is the elimination of the web server in between the fuzzer and the web application in order to achieve higher performance.
While such performance improvements, which could be adopted into \phuzz~by future work, are an important research aspect, \phuzz~placed its focus on supporting a great variety of client-side and server-side vulnerability classes. 
Atropos supports 8 server-side and no client-side vulnerabilities, whereas \phuzz~is able to detect 2 classes of client-side vulnerabilities, a similar set of server-side vulnerabilities. While \phuzz~supports external entity injections (XXE), Atropos can detect dangerous file-uploads and server-side request forgery (SSRF). 
However, the latter could be detected by \phuzz~similar to the path traversals with straightforward engineering effort, e.g., by checking if URLs are passed to \texttt{file\_get\_contents()} or other vulnerable functions.
\phuzz'~already hooks PHP functions used for processing uploaded files, so supporting this vulnerability class is a question of engineering effort to modify the \textit{Fuzzer} component to support \texttt{multipart/form-data} HTTP requests. 

In contrast to \phuzz, Atropos only supports PHP version 7 due to hooking several low-level C-functions in the PHP interpreter. 
\phuzz~supports PHP7, the most recent version PHP8 and future PHP versions supported by the UOPZ extension, due to only hooking (non-)native PHP-language functions.

Further, Atropos implements snaphotting, e.g., to avoid side-effects in stateful applications from consecutive fuzz iterations such as triggering a logout.
Although \phuzz~offers function hooks for great flexibility in changing the target application's behavior, extending \phuzz~with snapshotting capabilities could be interesting to future work. 

\subsubsection{Witcher}
Another recent work in this research area is \cite{trickel2022toss}, a greybox coverage-guided fuzzer for SQL injection and command injection vulnerabilities for several languages.
In fact, the Witcher was published concurrent to the work on \phuzz.
Both their and our work identify similar challenges for applying coverage-guided fuzzing to web applications. 

Witcher extends the concept of \emph{fault escalation} by instrumenting external parsers, i.e., from the database or web server, and checking for parsing errors to identify vulnerabilities.
This requires hooks into \texttt{libc}'s \texttt{recv} function using \texttt{LD\_PRELOAD} of the database process for SQL injection and replacing \texttt{/bin/sh} with a custom modified version of Dash for command injection detection.
While applying binary-level hooks to the interpreters allows Witcher to fuzz web applications written in different languages, including PHP, it has several limitations that do not apply to \phuzz.

For one, the fault escalation turns parsing errors into segmentation faults, thereby terminating the interpreter when encountering the first vulnerability to allow the fuzzer to detect it.
With \phuzz, we have noticed that terminating on the discovery of the first vulnerability is not always desirable, as it masks any other potential vulnerabilities in that code path; a phenomena called bug shadowing \cite{ulitzsch2023asanity}.
Some of the vulnerabilities in Table \ref{tab:fuzz-known-vulns-wp} were only discovered because \phuzz~did not terminate on the first finding.
Second, requiring low-level modification of external components for vulnerability detection is additional work, thus raising the bar to entry for developers, as well as increasing the risk of errors and inaccuracy, especially when the communication protocol is not plaintext or simple.

Although \phuzz~is limited to PHP, all instrumentation happens centrally at the interpreter, allowing \phuzz~to support any external component whose potentially vulnerable PHP functions were hooked.
Lastly, Witcher's approach limits the vulnerability classes it can detect, while \phuzz~is limited by the feasibility of hooking potentially dangerous PHP functions, enabling \phuzz~to support more vulnerability classes.
\subsubsection{CeFuzz}
This fuzzer combined static and dynamic analysis of a PHP web application to detect RCE vulnerabilities and was presented in \cite{Zhao2022}.
The approach first performs a static analysis of the target's source code to determine possible vulnerabilities. 
Then, the source code is instrumented with checkpoints, requiring changes to the source code, i.e., additional output statements to provide feedback to the fuzzer.
Although these modifications are claimed to be harmless, placing output statements in a target's source code can have unknown side effects, especially when the target uses output buffering to capture its output. 
Further, the vulnerability detection is based on the output or side effects of successfully executing specific commands.
However, not all RCE vulnerabilities allow complete control over what is executed, limiting the detection accuracy of this approach.
Lastly, CeFuzz appears to be closed source, so no comparison to \phuzz~is possible.
\phuzz~overcomes CeFuzz' limitations by instrumenting on the PHP interpreter level and detecting RCE vulnerabilities based on invalid commands passed to the underlying shell, not requiring a successful execution. 
\subsubsection{webFuzz}
This work only focuses on detecting cross-site scripting vulnerabilities in PHP web applications with instrumentation on the AST level \cite{van2021webfuzz}.
In contrast to \phuzz, the instrumentation requires actively changing the target's source code, which bears the risk of introducing errors:
First, webFuzz inserts several lines of specific PHP code at the beginning of \emph{every} PHP file to initialize the coverage collection.
Second, the additional coverage collection statements are inserted into the PHP file at various points with a transformation to an AST and back. 
Although webFuzz' coverage guidance is closer to AFL's implementation, \phuzz' line-based coverage provided by Xdebug or PCOV essentially covers the same basic blocks.
Nonetheless, webFuzz is open source, and its XSS detection is adopted into \phuzz.

\subsubsection{Other}
Greybox fuzzing, as performed by \phuzz, is not the only fuzzing strategy suitable for finding vulnerabilities in web applications.
Many blackbox fuzzers exist, of either commercial (\cite{burp_2023,acunetix_2023}), open source (\cite{w3af_2023,sqlmap_2023,zap_2023,wapiti,wfuzz}), or academic (\cite{Huang2021,Lee2020}) nature. 
While this allows these fuzzers to be used against any web application, i.e., whose source code is unavailable, their vulnerability detection is based on limited feedback that is observable from the outside.
PHP's error and exception handling and web application firewalls or input transformation can render this approach ineffective.
\phuzz~overcomes these limitations using its function hooking and interpreter-level instrumentation to detect otherwise unobservable vulnerabilities.

On the opposite side are whitebox fuzzers that use static analysis methods, such as taint analysis or symbolic execution, to identify vulnerabilities \cite{Jovanovic2006,Dahse2014,Alhuzali2016,alhuzali2018navex}.
These approaches require access to the source code and face different limitations.
For example, Pixy, Chainsaw, and Navex lack support for specific PHP features, limiting the applicability of these tools to modern web applications.
Although \phuzz~also requires the source code, it supports the most recent PHP version, and its runtime analysis can provide additional information to identify vulnerabilities, such as runtime exceptions, error, and stack traces, as explained in section \ref{sec:eval:known}.

\subsection{Future Work}
We mentioned in earlier sections that \phuzz~only implements basic and unsophisticated algorithms for vulnerability detection, candidate mutation, and candidate selection.
Future work could explore adopting new algorithms or optimizations that proved successful in other fuzzers to improve \phuzz.
The modularity and object-orientedness of \phuzz~should facilitate new research in that field.

Extending UOPZ' hooking functionality to PHP expressions could help to support more vulnerabilities classes and better identify vulnerabilities, e.g., file inclusions and code injection.

While \phuzz~already supports more client-side and server-side vulnerability classes than any other academic coverage-guided fuzzer, additional vulnerability classes, e.g., server-side request forgery (SSRF), dangerous file-uploads, HTTP header injection, or others, could be implemented with additional function hooks and \textit{VulnCheckers}.

\subsection{Ethical Considerations  \& Open Science}
As discussed in section \ref{sec:zero-day-vulns}, the authors followed responsible disclosure principles to notify the affected plugin developers about the discovered security issues. 
Furthermore, CVE-IDs were requested at the Certified Numbering Authority responsible for WordPress and its plugins.
All vulnerability details will be published upon remediation by the affected plugin developers or after a suitable grace-period, as the risk of in-the-wild exploitation is considered low due to the authentication and permission requirements.

In order to enable future work and the reproducibility of this work, we open source \phuzz~and all related code or datasets on Github \cite{phuzz_repo}. 

\section{Conclusion}
This work discussed several challenges of applying coverage-guided fuzz testing to PHP web applications.
Our proposed solution is \phuzz, a modular framework implementing novel approaches to tackle the identified challenges.
For example, \phuzz'~instrumentation neither requires changes to the fuzzed application's or PHP interpreter's source code nor external components, something not achieved by related work before.

Further,  we have evaluated \phuzz~against a large and diverse set of artificial and real-world PHP web applications, testing for known and unknown vulnerabilities and comparing it to several state-of-the-art fuzzers.
The results show that our instrumentation and vulnerability detection approach does not negatively affect the fuzzed applications while outperforming related work in the number of detected vulnerabilities and supported client-side and server-side vulnerabilities classes.
In fact, we discovered over 20 security issues, including two new zero-day vulnerabilities, in authenticated APIs by fuzzing 115 of the most popular WordPress plugins, showcasing \phuzz' effectiveness.

After all, we hope that making the modular framework openly accessible lays the foundation for future work in this research field and possible improvements to \phuzz. 
\vfill




%
\pagebreak
\bibliographystyle{ACM-Reference-Format}
\bibliography{bibliography}

\pagebreak
\appendix

\section{Appendix}\label{sec:appendix}

\begin{table}[h]
	\centering
	\caption{Hooked (non-)native PHP functions to detect vulnerabilities and to disable authentication and authorization checks}\label{tab:phuzz-uopz-hooks}
	\begin{tabular}{lp{0.8\linewidth}}
        \toprule
		{Type}  & {PHP functions} \\ \midrule
		SQLi & \texttt{mysqli\_query}, \texttt{mysqli::query}, \texttt{PDO::query}, \texttt{PDO::exec} \\
		XXE & \texttt{DOMDocument::load}, \texttt{DOMDocument::loadXML} \\
		RCE & \texttt{shell\_exec}, \texttt{system}, \texttt{passthru}, \texttt{exec} \\
		IDes & \texttt{unserialize} \\
		PaTr & \texttt{chgrp}, \texttt{chmod}, \texttt{chown}, \texttt{clearstatcache}, \texttt{copy}, \texttt{disk\_free\_space}, \texttt{disk\_total\_space}, \texttt{file\_exists}, \texttt{file\_get\_contents}, \texttt{file\_put\_contents}, \texttt{file}, \texttt{fileatime}, \texttt{filectime}, \texttt{filegroup}, \texttt{fileinode}, \texttt{filemtime}, \texttt{fileowner}, \texttt{fileperms}, \texttt{filesize}, \texttt{filetype}, \texttt{fnmatch}, \texttt{fopen}, \texttt{is\_dir}, \texttt{is\_executable}, \texttt{is\_file}, \texttt{is\_link}, \texttt{is\_readable}, \texttt{is\_uploaded\_file}, \texttt{is\_writable}, \texttt{lchgrp}, \texttt{lchown}, \texttt{link}, \texttt{linkinfo}, \texttt{lstat}, \texttt{mkdir}, \texttt{move\_uploaded\_file}, \texttt{parse\_ini\_file}, \texttt{parse\_ini\_string}, \texttt{readfile}, \texttt{readlink}, \texttt{realpath}, \texttt{rename}, \texttt{rmdir}, \texttt{stat}, \texttt{symlink}, \texttt{tempnam}, \texttt{touch}, \texttt{unlink} \\
		WordPress & \texttt{check\_admin\_referer}, \texttt{is\_admin}, \texttt{check\_ajax\_referer}, \texttt{current\_user\_can}, \texttt{get\_current\_user\_id}, \texttt{get\_user\_meta}, \texttt{is\_super\_admin}, \texttt{is\_user\_logged\_in}, \texttt{user\_can}, \texttt{wp\_get\_current\_user}, \texttt{wp\_verify\_nonce} \\
        \bottomrule
	\end{tabular}
\end{table}

\begin{lstlisting}[caption={Definition of API endpoints and handlers in a WordPress plugin.},style=php,label={lst:wpapi}]
    add_action('wp_ajax_myfunc','myfunc');
    add_action('wp_ajax_nopriv_myapi','myapi');
    function myfunc() {/*code*/}
    function myapi() {/*code*/}
\end{lstlisting}

\begin{lstfloat}[b!]
\begin{lstlisting}[caption={Example HTTP request (lines 1-5) and HTTP response (lines -14)},label={lst:examplehttpreq},style=nocoloring]
	GET /page.php?id=123 HTTP/1.1
	Host: example.org
	Cookie: PHPSESSID=onprnes23regokc[...]
	Connection: close
	
	
	HTTP/1.1 200 OK
	Date: Fri, 20 Jan 2023 12:11:36 GMT
	Server: Apache/2.4.54 (Debian)
	X-Powered-By: PHP/8.1.14
	Content-Length: 642
	Content-Type: text/html;charset=utf-8
	
	<html>[...]
\end{lstlisting}
\end{lstfloat}

\begin{figure}[ht]
    \centering
    \includegraphics[width=1\linewidth]{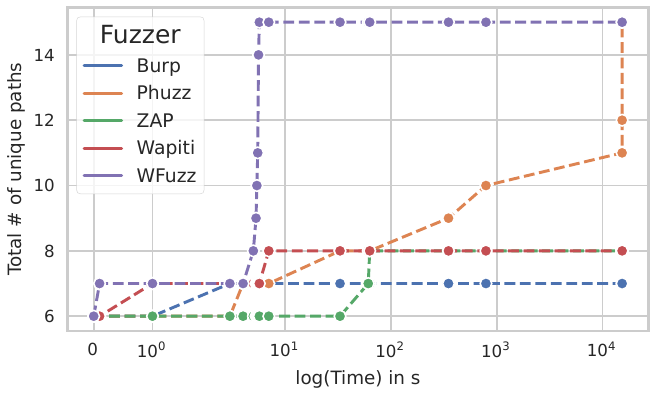}
    \caption{\phuzz~continuously finds new coverage and eventually the vulnerability in our custom DVWA \emph{fuzz} SQLi level. WFuzz only reaches the same coverage accidentally by having the right keyword in its wordlist, but does not identify the vulnerability.}
    \Description{A line plot showing the amount of coverage discovered over time for \phuzz, BurpSuite, ZAP, Wapiti and WFuzz for the special DVWA SQLi \emph{fuzz} level.}
    \label{fig:phuzzvsburp}
\end{figure}

\begin{lstlisting}[style=php,caption={Example \phuzz~configuration file for the command injection level with difficulty \emph{high} in DVWA},label={lst:fuzzconfig}]
	{
		"target":"http://web/vulnerabilities/exec/",
		"login": "dvwa_requests",
		"methods": [
		    "POST"
		],
		"cookies": {
			"data": [
			{
				"name": "security",
				"value": "high"
			}
			],
			"fixed": [
			    ".*"
			],
			"fuzz": [],
			"login": [
			    "PHPSESSID"
			],
			"weight": 0.0
		},
		"body_params": {
			"data": [
			{
				"name": "ip",
				"seeds": [
				    "fuzz"
				]
			},
			{
				"name": "Submit",
				"value": "Submit"
			}
			],
			"fixed": [
			    "Submit"
			],
			"fuzz": [
			    ".*"
			],
			"weight": 1.0
		}
	}
\end{lstlisting}

\begin{table}[ht]
	\centering
	\caption{Fuzzed WordPress plugins with known vulnerabilities}\label{tab:phuzz-known-vulns-wp-ap}
	\begin{tabular}{lp{0.3\linewidth}p{0.4\linewidth}}
        \toprule
		{Vuln. class}  & {Plugin name} & {Vuln. report's title} \\ \midrule
		XSS & show-all-comments-in-one-page & $<$ 7.0.1 - Reflected XSS \\ 
		XSS & essential-real-estate & $<$ 3.9.6 - Reflected Cross-Site-Scripting\\  
		XSS & crm-perks-forms & $<$ 1.1.1 - Reflected XSS\\  
		XSS & rezgo & $<$ 4.1.8 - Reflected Cross-Site-Scripting\\  
		XSS & gallery-album & $<$ 2.0.0 - Reflected Cross-Site Scripting\\  
		SQLi & 	kivicare-clinic-management-system& $<$ 2.3.9 - Unauthenticated SQLi\\  
		SQLi & nirweb-support & $<$ 2.8.2 - Unauthenticated SQLi\\  
		SQLi & 	arprice-responsive-pricing-table& $<$ 3.6.1 - Unauthenticated SQLi\\  
		SQLi & ubigeo-peru & $<$ 3.6.4 - Unauthenticated SQLi\\  
		SQLi & 	photo-gallery & $<$ 1.6.3 - Unauthenticated SQL Injection\\  
		PaTr & usc-e-shop & $<$ 2.8.5 - Unauthenticated Arbitrary File Access\\  
		PaTr &udraw & $<$ 3.3.3 - Unauthenticated Arbitrary File Access\\  
		PaTr & seo-local-rank& $<$ 2.2.4 - Unauthenticated Arbitrary File Access via Path Traversal\\  
		PaTr &	hypercomments & $<=$ 1.2.2 - Unauthenticated Arbitrary File Deletion\\  
		PaTr & nmedia-user-file-uploader& $<$ 21.3 - Unauthenticated File Renaming\\  
		OpRe&newsletter-optin-box & $<$ 1.6.5 - Open Redirect\\  
		OpRe&webp-converter-for-media & $<$ 4.0.3 - Unauthenticated Open redirect\\  
		OpRe& phastpress& $<$ 1.111 - Open Redirect\\  
		OpRe& 	pie-register& $<$ 3.7.2.4 - Open Redirect\\  
		OpRe& all-in-one-wp-security-and-firewall& $<=$ 4.4.1 - Open Redirect \& Hidden Login Page Exposure\\ 
		IDes & totop-link& $<=$ 1.7.1 - Unauthenticated PHP Object Injection\\
		IDes &	joomsport-sports-league-results-management & $<$ 5.1.8 - Unauthenticated PHP Object Injection\\
        \bottomrule
	\end{tabular}
\end{table}

\begin{table*}
    \small
	\centering
	\caption{Results for DVWA by vulnerability level and difficulty}\label{tab:fuzz-dvwa}
	\begin{tabular}{lcccc}
		\toprule
		{Difficulty:} & \shortstack{{\phuzz}\\(low/medium/high)} & \shortstack{{BurpSuite}\\(low/medium/high)} & \shortstack{{ZAP}\\(low/medium/high)} & \shortstack{{Wapiti}\\(low/medium/high)} \\ \midrule
		RCE & \ok~/ \ok~/ \ok & \ok~/ \ok~/ \ok & \ok~/ \nok~/ \nok & \ok~/ \ok~/ \ok \\
		SQLi & \ok~/ \ok~/ (\ok)$^2$ & \ok~/ \ok~/ -$^2$ & \ok~/ \ok~/ -$^2$ & \ok~/ \ok~/ -$^2$ \\
		SQLi blind & \ok~/ \ok~/ \ok & \ok~/ \nok~/ \ok & \ok~/ \nok~/ \nok & \nok~/ \nok~/ \nok\\
		XSS reflected & \ok~/ \ok~/ \ok & \ok~/ \ok~/ \ok & \ok~/ \ok~/ \ok& \ok~/ \ok~/ \ok\\
		XSS stored & \ok~/ \ok~/ \ok & \ok~/ \ok~/ \ok &\ok~/ \ok~/ \ok& \ok~/ \ok~/ \ok\\
		File inclusion & \ok$^1$~/ \ok$^1$~/ \ok & \ok~/ \ok~/ \nok &\ok~/ \ok~/ \nok& \ok~/ \ok~/ \ok\\ 
		\bottomrule
	\end{tabular}
	\\ \ok~Found, \nok~Not found, $^1$~Indicated by errors, $^2$~Requires multi-step support
\end{table*}

\begin{table*}
    \small
	\centering
	\caption{Results for bWAPP by vulnerability class}\label{tab:fuzz-bwapp}
	\begin{tabular}{lcccc}
		\toprule
		{Vuln. class} ($N$) & {\phuzz} & {BurpSuite} & ZAP & Wapiti \\ \midrule
		\shortstack{XSS (14)\\~\\~} & \shortstack{\ok~/ \ok~/ \ok~/ \ok~/ \ok~/\\ \ok~/ \ok~/ \ok~/ \ok~/ \ok~/\\ \ok~/ \ok~/ \ok~/ \ok} & \shortstack{\ok~/ \ok~/ \ok~/ \ok~/ \ok~/\\ \ok~/ \ok~/ \ok~/ \nok~/ \ok~/\\ \ok~/ \ok~/ \ok~/ \ok} & \shortstack{\ok~/ \ok~/ \nok~/ \ok~/ \ok~/\\ \ok~/ \ok~/ \nok~/ \nok~/ \ok~/\\ \ok~/ \nok~/ \nok~/ \nok} & \shortstack{\ok~/ \ok~/ \ok~/ \ok~/ \ok~/\\ \ok~/ \ok~/ \nok~/ \nok~/ \ok~/\\ \ok~/ \nok~/ \nok~/ \ok} \\
		\shortstack{SQLi (9)\\~} & \shortstack{\ok~/ \ok~/ \ok~/ \ok~/ \ok~/\\ \ok~/ \ok~/ \ok~/ \ok} & \shortstack{\ok~/ \ok~/ \ok~/ \ok~/ \ok~/\\ \ok~/ \ok~/ \ok~/ \ok} & \shortstack{\ok~/ \ok~/ \ok~/ \ok~/  \ok~/\\ \ok~/ \ok~/ \ok~/ \ok}& \shortstack{\ok~/ \ok~/ \ok~/ \ok~/  \nok~/\\ \ok~/ \ok~/ \ok~/ \nok~}\\
		RCE (3) & \ok~/ \ok~/ \ok$^1$ & \ok~/ \ok~/ \nok & \timeout~/ \ok~/ \ok & \timeout~/ \timeout~/ \nok\\
		PaTr (2) &\ok$^1$~/ \ok & \ok~/ \ok & \ok~/ \ok & \ok~/ \ok\\
		OpRe (2) & \ok~/ \ok& \ok~/ \ok & \ok~/ \ok & \ok~/ \ok\\
		\bottomrule
	\end{tabular}
	\\ \ok~Found, \nok~Not found, $^1$~Indicated by errors, $^2$~Requires multi-step support, \timeout~Timeout
\end{table*}

\begin{table*}
	\centering
	\caption{Results for XVWA by vulnerability class}\label{tab:fuzz-xvwa}
	\begin{tabular}{lcccc}
		\toprule
		{Vuln. class} ($N$) & {\phuzz} & {BurpSuite} & ZAP & Wapiti \\ \midrule
		SQLi (4) & \ok~/ \ok~/ \ok~/ \ok & \nok~/ \ok~/ \nok~/ \ok & \nok~/ \nok~/ \nok~/ \ok& \nok~/ \nok~/ \nok~/ \nok \\
		XSS (2) & \ok~/ \ok & \ok~/ \nok & \ok~/ \timeout & \ok~/ \ok\\
		RCE (1) & \ok & \ok & \ok & \ok \\
		IDes (1) & \ok & \nok & \nok & \nok \\
		PaTr (1) & \ok$^1$ & \ok & \ok & \ok \\
		OpRe (1) & \ok & \ok & \ok & \ok \\  
		\bottomrule
	\end{tabular}
	\\ \ok~Found, \nok~Not found, $^1$~Indicated by errors, $^2$~Requires multi-step support, \timeout~Timeout
\end{table*}

\begin{table*}
	\centering
	\caption{Results for WackoPicko by vulnerability class}\label{tab:fuzz-wackopicko}
	\begin{tabular}{lcccc}
		\toprule
		{Vuln. class ($N$)} & {\phuzz} & {BurpSuite} & ZAP & Wapiti \\ \midrule
		XSS (4) & \ok~/ \ok~/ \ok~/ \ok & \ok~/ \ok~/ \ok~/ \ok & \ok~/ \ok~/ \ok~/ \ok& \ok~/ \ok~/ \ok~/ \ok\\
		RCE (1) &  \ok &  \nok & \timeout & \timeout \\
		SQLi (1) & \ok &  \ok  & \ok & \nok \\
		PaTr (1) & \ok$^1$  & \nok  & \ok & \ok \\ 		
		\bottomrule
	\end{tabular}
	\\ \ok~Found, \nok~Not found, $^1$~Indicated by errors, $^2$~Requires multi-step support, \timeout~Timeout
\end{table*}

\begin{table*}
	\centering
	\caption{Results for WordPress plugins by vulnerability class}\label{tab:fuzz-known-vulns-wp}
	\begin{tabular}{lcccc}
		\toprule
		{Vuln. class} ($N$) & {\phuzz} & {BurpSuite} & ZAP & Wapiti \\ \midrule
		XSS (5) & \ok~\ok~\ok~\ok~\ok & \ok~\ok~\ok~\ok~\ok & \ok~/ \timeout~/ \ok~/ \ok~/ \timeout & \ok~/ \timeout~/ \ok~/ \ok~/ \timeout \\
		SQLi (5) & \ok~\ok~\ok~\ok~\ok & \nok~\nok~\nok~\nok~\nok & \timeout~/ \timeout~/ \timeout~/ \timeout~/ \timeout~/ & \timeout~/ \nok~/ \nok~/ \nok~/ \nok~/ \\
		PaTr (5) & \ok~\ok~\nok~\ok~\ok & \ok~\nok~\nok~\nok~\nok & \ok~/ \timeout~/ \nok~/ \timeout~/ \timeout& \ok~/ \timeout~/ \nok~/ \timeout~/ \nok\\
		OpRe (5) & \ok~\ok~\ok~\ok~\ok & \ok~\ok~\ok~\ok~\ok & \timeout~/ \ok~/ \ok~/ \ok~/ \ok & \ok~/ \ok~/ \ok~/ \ok~/ \ok\\ 
		IDes (2) & \ok$^1$~\ok$^1$ & \nok~\nok & \nok~/ \timeout& \nok~/ \nok \\ 
		\bottomrule
	\end{tabular}
	\\ \ok~Found, \nok~Not found, $^1$~Indicated by errors, $^2$~Requires multi-step support, \timeout~Timeout
\end{table*}

\end{document}